# TO THE ACCELERATION OF CHARGED PARTICLES WITH TRAVELLING LASER FOCUS

A. Mikhailichenko[1], Cornell University, CLASSE, Ithaca, NY14853, USA

*Abstract.* We describe here the latest results of calculations with FlexPDE code of wake-fields induced by the bunch in micro-structures. These structures, illuminated by swept laser bust, serve for acceleration of charged particles. The basis of the scheme is a fast sweeping device for the laser bunch. After sweeping, the laser bunch has a slope ~45º with respect to the direction of propagation. So the every cell of the microstructure becomes excited locally only for the moment when the particles are there. Self-consistent parameters of collider based on this idea allow consideration this type of collider as a candidate for the near-future accelerator era.

## OVERVIEW

Future of High-Energy physics in a post-Standard Model world associated with implementation of new technologies suitable for particles acceleration. Main criterions are the rate of acceleration and luminosity. In [1-2,9-12] we made description of method of acceleration with local excitation of microstructure by swept laser radiation. This method, based on idea disclosed in [3], was called a *Travelling Laser Focus* (TLF).

Acceleration in any structure is going by interaction of particles with standing in transverse direction wave, so basically this is an interaction with *two* waves propagating oppositely in a transverse direction; this allows exclusion of magnetic field action while the particle crosses the center of (micro) cavity[2]. Using the quantum mechanical language one can say that acceleration happens thanks to interaction of particle with *two* photons synchronously moving in a transverse direction. So the presence of two photons (standing wave) is important in acceleration process. Otherwise the scattering of particle on accelerating EM wave reduces achievable rate by Compton-effect, which immediately manifests itself, due to the high density of beam and laser radiation. That is why acceleration with the only single tilted laser bunch like it is described in [4] is not working for high-energy acceleration.

As a baseline we have chosen a foxhole type of structure excited from the open side, introduced in [5] and developed further in [6]. In our case the standing wave arranged inside each tiny accelerating cell of foxhole-type structure by multiple reflections of laser wave inside each cell, pretty much in the same way as in an ordinary RF cavity. Dimensions of structure are chosen of the order of the wavelength of laser radiation, which is $\lambda_{ac}$ ~1$\mu m$.

Principle of excitation of accelerator structure is represented in Fig.1, where the photon bunch with the length $l_b$~$c\tau$~$3cm$ tilted transversely in direction of particle motion by the angle ~45º so the focal point following the micro-bunch of relativistic particles. Longitudinal size of the laser spot on the structure is defined by the method of illumination used. The transverse size defined by a short-focusing cylindrical lens, installed in front of the accelerating structure. For producing of tilted laser bunch we appointed a method which was developed earlier, see [2], which uses the fast sweeping device arranged by many electro-optical prisms with reversing orientation of optical axes. These prisms located between plates of matched strip-line. This strip line excited with EM pulse co-propagating with the laser bunch.

---

[1] aam10@cornell.edu.
[2] For cylindrical cavity standing wave in radial direction described by Bessel functions.

Other way to generate the tiled laser bunch –with help of optical diffraction grating [16,17] – is much less productive for high-energy acceleration, see below.

## THE CONCEPT

The concept of feeding accelerator structure with sloped laser pulse is represented in Fig.1. Here, the accelerating structure is a $2\pi$ type as the period of structure is equal to the wavelength of laser radiation $\lambda_{ac}$. The EM field phases in neighboring cells are the same. Graphs of $E(t)$ in Fig.1 represent the time dependence of electric field in the gaps of two cells at different locations. The particle meets an accelerating phase in a cell's gap *inside* the structure and decelerating phase it meets in the narrowing between the cells where the field is exponentially smaller. This is similar to an ordinary accelerating structure with the drift tubes. So there is no slippage between particle and accelerating field by definition. Slope (tilt) of the laser bunch $\alpha$ defined by the speed of particle, $\tan\alpha = v/c$, so $\alpha = 45°$ for relativistic particles. The slope could have deviations from the linear one, as this affects the average accelerating gradient only [2, 3]. Illumination lasts for the time $\tau \cong l_t/c$, $l_f \approx l_t$, Fg.1, while the laser bunch length is $c\tau \cong 3\ cm$. Each cell of structure illuminated by the time defined by the width of the laser spot $l_t$, Fig.1. The width $l_f$ defines how many micro-bunches could be in a train of accelerated bunch.

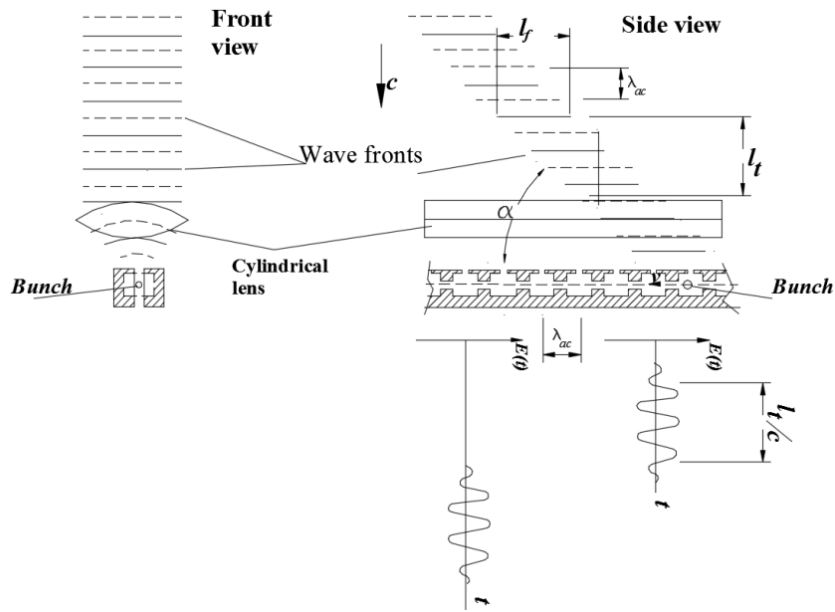

Figure 1: The concept of TLF illumination [3]. The time dependence of the field $E(t)$ at two different locations along the structure is represented on the graphs at the bottom. Bunch has a velocity directed to the left.

# TECHNIQUES FOR GENERATION OF SLOPED LASER BUNCH

We consider here briefly two ways on how to make sloped laser bunch with the wave fronts directed perpendicular to direction of its propagation (see detailed description in [2,10]).

## *Sweeping of laser pulse*

Each sweeping device uses controllable deflection of the laser radiation in time. The device could be characterized by the deflection angle $\vartheta$ and the angle of natural diffraction $\vartheta_d \cong \lambda_{ac}/a$, where $a$ —is the aperture of the sweeping device. The ratio of deflection angle to diffraction angle is a fundamental measure of the quality for any deflecting device. This ratio defines the *number of resolved spots* (pixels) on the surface of the structure, $N_R = \vartheta/\vartheta_d$. The deflection angle could be increased by appropriate optics, but the number of resolved spots $N_R$ is an invariant under such transformations. The value of $N_R$ gives the number for the lowering of the laser power required for gradient desired and, also, the number for the duty of illumination time reduction. The last is important for lowering of structure heating. So it is desirable to have this number as big as possible.

In this case the length of laser bunch is about the length of the accelerating structure, Fig.2. The lens installed in front of the sweeping device has the focal point at the structure. As we already mentioned, deflection angle could be changed by appropriate optics but the ratio of deflecting angle $\vartheta_{sw}$ to the diffraction angle– $N_R = \vartheta_{sw}/\vartheta_d^s \sim 100$ –remains the same.

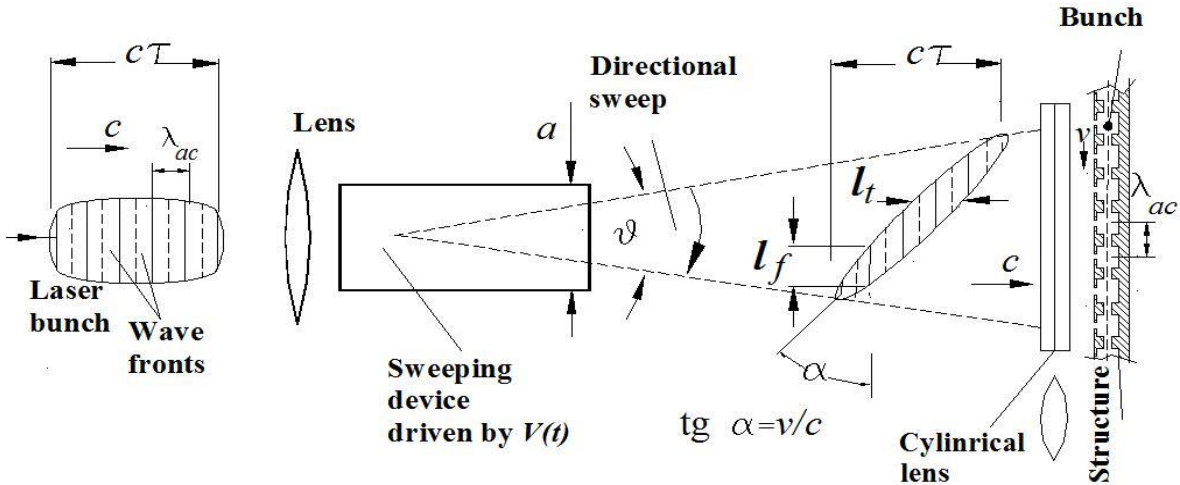

Figure 2: TLF principle of preparation of sloped laser bunch with the sweeping device [3]. Electron bunch is moving from the top to the bottom inside the structure.

For sweeping of ~3*cm*-long laser bunch we appointed a multi-prism device feed with *traveling* EM pulse, suggested in [2]. This EM pulse has a slope along the bunch ~few *kV* arranged either by the pulser, synchronized by the laser pulse itself or by a *cm*-wavelength RF. In the last case the laser pulse appearance is synchronized with RF phase. Few examples of engineering of the sweeping device can be found in [10].

*Sweeping device with electro-optical crystals.*

For a prism-based device, Fig.3, change in refraction index yields the change in deflection angle. To arrange such a change, the basements of the prism must be covered by metallic foils and a high voltage applied to them.

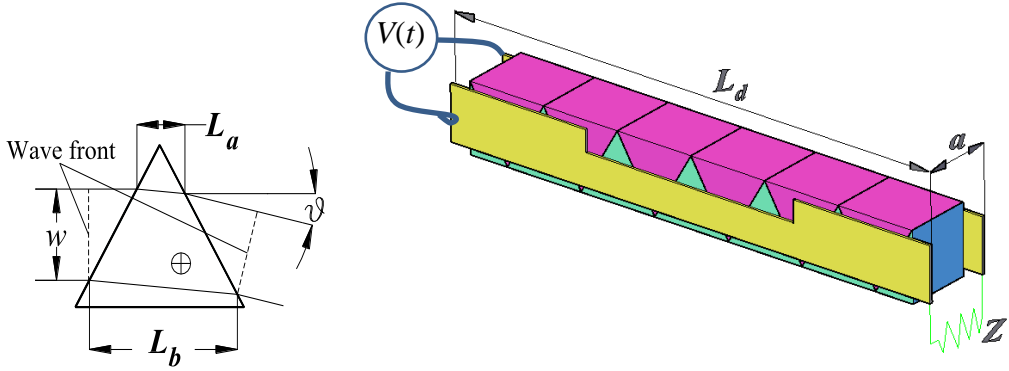

Figure 3: The prism deflection device concept, left. Cross marks direction of optical axis. At the right- the prisms with *oppositely directed optical axes* installed in series between two parallel strip–line electrodes, Electromagnetic pulse $E_x(t - y/c)$ co-propagates with the laser bunch to the right.

For a prism-based device, Fig.3, left, the deflecting angle is defined by the phase delay across the laser beam front arising from differences in the path lengths in material of the prism having a refractive index $n$

$$\vartheta \cong n\frac{(L_a - L_b)}{w}, \qquad (1)$$

where $w \sim a$ –is the width of incident laser beam, $L_a$ and $L_b$ –are the distances through which the edges of the laser beam traverse the prism, Fig.3. A change in refractive index value yields a deflection angle change. If the basements of the prism are covered by metallization, then voltage $V(t)$ applied to the metallization will change refractive index as $\Delta n = \Delta n(V(t))$, so

$$\Delta\vartheta(t) \cong \Delta n(t)\frac{(L_a - L_b)}{w}. \qquad (2)$$

The number of resolvable sports for this device $N_R$ can be found now as

$$N_R \cong \frac{|\Delta\vartheta|}{\lambda_{ac}/w} = \frac{\Delta n \cdot |L_a - L_b|}{\lambda_{ac}} \cong \Delta n \frac{l}{\lambda_{ac}}, \qquad (3)$$

where $l = |L_a - L_b|$ stands for the prism base length and $a/w \cong 1$ in our case. In any case, a shorter wavelength is preferable from this point of view.

One can see from (1) and (2) that with increase in the optic path difference $l$, both the deflection angle and the number of resolved spots increase also. To increase the last numbers multiple-*prism* deflectors were developed, see Fig.3, right. Here the neighboring prismatic crystals have oppositely oriented optical axes. These crystals positioned between the strip–line electrodes. In this case the full length of a sweeping device $L_d$ serves as $l$ in formulas (2), (3). To be able to sweep short laser bunches, the voltage pulse $V(t)$ is propagating along this strip-line as a *traveling wave* together with the laser bunch to be swept, as it was proposed in [2,3]. This gives the necessary voltage profile along the laser pulse at *cm* distances, what corresponds to the pulse duration $c\tau$, which is ~100*ps* typically [2,3].

For Cartesian coordinate system coinciding with the principal axes of a crystal, the indicatrix could be represented as

$$\left(\frac{1}{n^2}\right)_1 x^2 + \left(\frac{1}{n^2}\right)_2 y^2 + \left(\frac{1}{n^2}\right)_3 s^2 + 2\cdot\left(\frac{1}{n^2}\right)_4 ys + 2\cdot\left(\frac{1}{n^2}\right)_5 xs + 2\cdot\left(\frac{1}{n^2}\right)_6 xy = 1. \qquad (4)$$

The refractive index in active media in presence of electric field has dependence like [2]

$$1/n_i^2 = 1/n_{0i}^2 + \sum_j r_{ij}\cdot E^j, \qquad (5)$$

where $r_{ij}$ –are components of a $6\times 3$ tensor, $n_0$ is a refraction index without field[3]. Index $i$ runs from 1 to 6; 1 stands for *xx*, 2–for *yy*, 3– *ss*, 4– for *ys*, 5–for *xs*, 6– for *xy*, *s*-longitudinal coordinate. Indicatrix allows to determinate the refractive index *n* components for monochromatic plane waves as a function of theirs polarization. For *KDP* tetragonal structure the principal components are $r_{41}=r_{52}\cong 8.8\cdot 10^{-12} m/V$; $r_{63}\cong 10.5\cdot 10^{-12} m/V$. The change of refractive index is equal to $\Delta n_i \cong (\partial n_i/\partial E_j)E^j(t)$. The (5) yields $\partial n_i/\partial E_j = -n_{0i}^3 r_{ij}/2$ and the net change of refractive index becomes

$$\Delta n_i \cong (\partial n_i/\partial E_j)E^j(t) \cong -n_{0j}^3 r_{ij} E^j/2. \qquad (6)$$

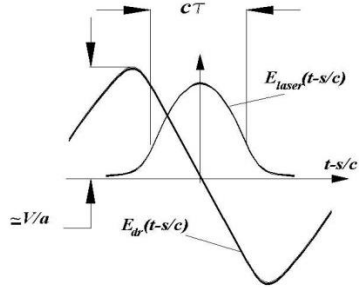

Figure 4: The laser pulse and EM driving wave are propagating in the same direction.

The deflection angle and the number of resolved spots become now

$$\Delta\vartheta \cong \Delta n(t)\frac{L_d}{w} \cong \frac{L_d}{a^2} n_0^3 \cdot r_{ij}\cdot V(t), \quad N_R \cong \frac{|\Delta\vartheta|_{max}}{\lambda_{ac}/w} = |\Delta n|_{max}\frac{2L_d}{\lambda_{ac}} = \frac{L_d}{\lambda_{ac}} n_0^3 r_{ij}|\Delta V|_{max}, \qquad (7)$$

where $L_d$ stands for the full length of deflecting device, *w*–is the laser beam width (along direction of deflection), $|\Delta\vartheta|_{max}$ stands for the full variation of deflection angle and the electrical field was substituted as $E=V/a$. For $L_d = 25 cm$, one can expect for $w\cong a\cong 0.5$ *cm*, that deflection angle becomes $\Delta\vartheta\cong 10^{-2}$ and $N_R \cong 20$ for $\lambda_{ac}\cong 10\mu m$ and, respectively $\Delta\vartheta\cong 10^{-2}$ and $N_R\cong 200$ for $\lambda_{ac}\cong 1\mu m$. We estimated the field amplitude applied to the crystals as 10 *kV/cm*. This is really a field strength *variation* along the laser bunch. This variation is traveling together with the laser beam along the sweeping device, Fig. 4.

---
[3] $n_0\approx 2.9$ for *ZnTe*, $n_0\approx 1.5$ @ $1\mu m$ for *KDP* (KH$_2$PO$_4$, Potassium dihydrogen Phosphate).

For acceleration, the laser flash energy passing through the single deflection system is ~10*mJ*. We estimated the area cowered by the laser beam as $S \cong w \cdot d$, what is about $0.25 cm^2$. This yields the energy density 40 $mJ/cm^2$ only. One can decrease the area at least 10 (up to 100) times. Reduction of orthogonal dimension will reduce the power required for deflection. One can install a final stage of the laser *after* sweeping device, so the power limitation will not be an issue here.

For such a sweeping device, a lot of electro-optical crystals can be used, see Table 1 below [2]. For example, a crystal *KDP* ($KH_2PO_4$) is transparent for a radiation with the wavelength $\lambda \cong 0.2 \div 1 \mu m$. Some other crystals, such as a *CdTe, CuCl, GaAs, ZnTe, ZnS* are transparent in the region of wavelengths around $\lambda \approx 10 \mu m$. The last group of materials have rather high refractive indexes $n_0$~(2-4) what compensate smaller electro- optical coefficient.

Table 1: Electro-Optical Materials for Deflector

| Wavelength | Materials | $\vartheta$, rad | $N_R$ |
|---|---|---|---|
| $\lambda_{ac} \cong 10 \mu m$ | GaAs, ZnTe, ZnS, CdS, CdTe, CuCl | 0.01-0.02 | 20 |
| $\lambda_{ac} \cong 5 \mu m$ | LiNbO$_3$, LiTaO$_3$, CuCl | 0.01-0.02 | 40 |
| $\lambda_{ac} \cong 1 \mu m$ | KDP, DKDP, ADP, KDA, LiNbO$_3$ | 0.01-0.02 | 200 |

Despite the materials which are transparent for the longer wavelengths have lower value of $r_{ij}$ –components, they have higher values of refracting index insead, so the variation of refractive index becomes about the same.

***Embodiment of Sweeping Device with pulsed voltage source***

One example of technical realization of this traveling wave deflector as a strip-line is represented in Fig. 5. Here, basically, the strip-line electrodes 1 have prism crystals 3 with opposite optical axes orientation in between. HV impulse applied to one end of the strip-line trough connectors 4, then propagates to the other end and further absorbed by the matching resistor 2. Slots in electrodes 1 made for proper current distribution across the line taking into account that the wave front $\sim c\tau$ becomes comparable with the transverse size of the line.

In Fig.5 there is shown the lens, 5 giving longitudinal focus to the swept laser beam and the power splitter 6, which splits a fraction of laser radiation propagating along the accelerator. This engineering realization is very suitable for primary testing of this principle. Powering of this device can be arranged with the power generator using for its operation Inversely Recovered Diodes technique [13-15]. In these types of techniques, the energy stored in an inductor while the current running through it. While fast-interrupted, this current flow raises voltage on the load. For interruption of the current the Drift-Step Recovery Diodes (DSRDs) used successfully. In more detail, the semiconductor transition filled by the carriers during direct charge flow first. Then the second pulse with opposite polarity runs through the transition during the time determined by evacuation of carriers (electron-hole plasma) accumulated in transition region. Typical power supply of this class operates with repetition rate ~ 1*MHz* providing the rise time down to 50 *ps* and voltage up to 30 *kV*. Size of this device typically is $350 \times 150 \times 300$ $mm^3$ [13-15]. Other possibilities of triggering described in [10].

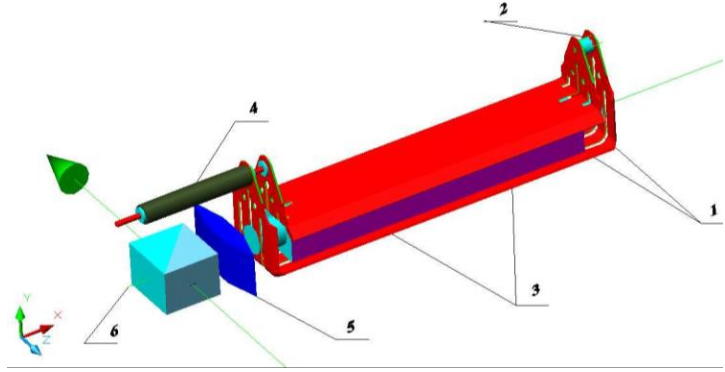

Figure 5: Example of technical realization of multi-prism sweeping device with a traveling wave and pulsed feeding. Electro-optical crystals 3 are positioned between the strip-line electrodes 1, loaded at the end with matching resistors 2. Excitation is going through the cable 4. 5 and 6 are the lens and splitting device respectively (see also [10]).

One can see that the speed of propagation of HV pulse along the stripline and the laser bunch should be kept equal. These could be arranged by placement of additional materials with dielectric permittivity and magnetic permeability outside the volume occupied by the prisms.

*Usage of diffracting grating for generation of sloped laser bunch.*

In this section we would like to discuss briefly the other method of creation the sloped laser bunch with the help of diffraction grating, see Fig. 6, [16,17]. First of all the laser bunch must be short in this method with time duty which is equal to the effective laser duty in previous method; this corresponds to $l_t$ in Fig.2. The pulse should have the width about twice the length of the accelerating structure which could be arranged with appropriate telescopic optics. In Fig.6 the incoming laser bunch hits the grating which has a tilt of $63^o$ with respect to direction of propagation. The effective length of the accelerating structure, which might be located at the bottom chosen the same as in previous section, $L \cong c\tau$. The geometric relations are clear from the Fig.6. As it follows from the principle of operation of grating, formation of reflection in necessary direction requires many periods and involves small area $\sim \delta$ on the grating only, Fig.6. This inevitably extends the laser pulse length. Really, the diffraction angle in this case is $\cong \lambda_{ac}/a$, where the area involved in formation of reflection chosen for comparison with the sweeping method as small as $\delta \approx l_f \cong l_t$.

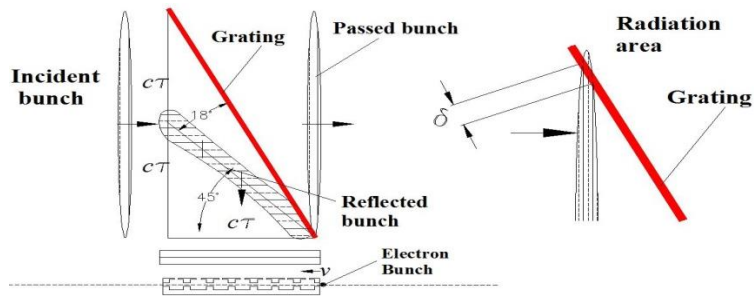

Figure 6: Geometry of changing the slope of laser bunch with (semi-transparent) grating, on the left. To the definition of effective radiation area-at the right.

For the sweeping device we have $l_t \cong L/N_R \cong a/N_R$. So for comparison of these two schemes, we represent the diffraction angle as $\vartheta_d^g \cong \sqrt{\lambda N_R/a}$. The ratio of diffraction angle in a grating to the diffraction one in a sweeping device in these two methods goes to be

$$\vartheta_d^g / \vartheta_d^s \cong \sqrt{\lambda_{ac} N_R / a}/(\lambda_{ac}/a) \cong \sqrt{N_R a/\lambda_{ac}} \quad . \tag{8}$$

With some optimization of grating profile this could be improved, probably, to $\vartheta_d^g / \vartheta_d^s \cong N_R$ at the best. So the advantage of using the sweeping device is obvious-it gives much smaller laser spot size in longitudinal direction.

The difference is ~100 times minimum in a favor of the sweeping device. Usage of sweeping device has undoubted advantages here [1].

## SCHEMES FOR LONG–TERM ACCELERATION

The scheme for long term acceleration with individual sweeping devices is represented in Fig. 7. The scheme in Fig.7 uses the optical splitting by semi-transparent mirrors. These mirrors split the laser bunch in fractions and direct each fraction to its individual accelerating structure.

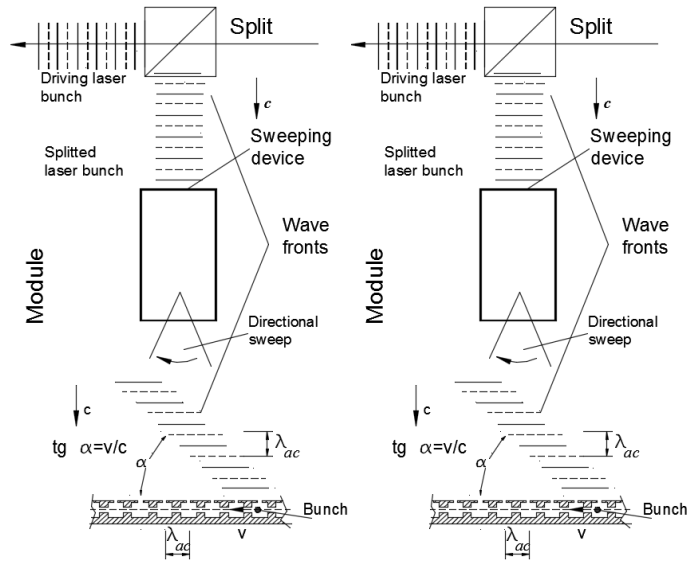

Figure 7: The concept of splitting the laser power for feeding a micro-structure. Two identical cells are shown.

The long term acceleration means the ability to accelerate low-emittance bunch of charged particles without its degradation. A very important component is the ability to accelerate the positively charged particles (positrons) in a view of arrangement of collisions at high energy.

Another realization of long-term acceleration with swept laser bunch is represented in Fig.8. One sweeping device can serve for few structures (~10). For compensation of difference in distances from the sweeping device to the structure, special lenses used here (marked as Correction lenses in Fig.8). These lenses together with the lenses installed around the sweeping device make location of laser focus on the particular structure.

We would like to underline here, that the sweeping device could be installed before the final stage of an optical amplifier. In this case the density of pumping optical energy could be reduced naturally. One can see, than this arrangement allows operation with the only single sweeping device as the optical amplifiers could be installed along the linac, while the laser power of tilted laser bunch reduced by the splitting devices.

The wave fronts in a laser bunch in Fig.8 are normal to the radial lines having theirs center in the sweeping device before the second focusing lens, which is located right after the sweeping device. This second lens has its focal point in a sweeping device.

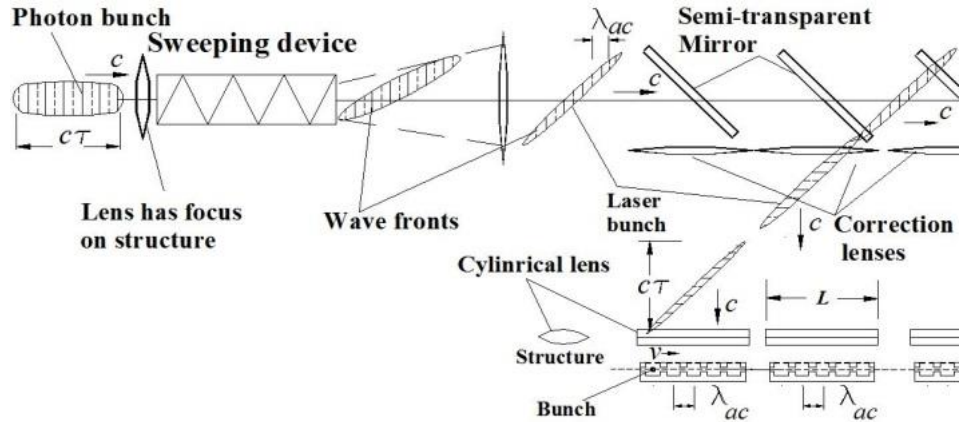

Figure 8: Long term acceleration system with the sweeping device [1]. Optical amplifier could be installed after the sweeping device (or in front of each cylindrical lens). Accelerated bunch is moving to the right.

We have chosen a foxhole type of structure excited from the open side as a baseline. This type of structure was introduced in [5, 6], although TLF method can work with any other appropriate structure type. We are suggesting a weak inductive coupling with the outer space, so the electric field at the open side is ~0, but magnetic field has a maximal value there. The height of structure should be then ~half of the wavelength in a groove-$\lambda_w/2$, where $\lambda_w \cong \lambda_{ac}/\sqrt{1-(\lambda_{ac}/2W)^2}/2$, $W$ is the width of cell, $\lambda_{ac}$ is a wavelength of laser radiation in free space, see [2]. Isometric 3D view of structure with cylindrical lens is represented in Fig.2

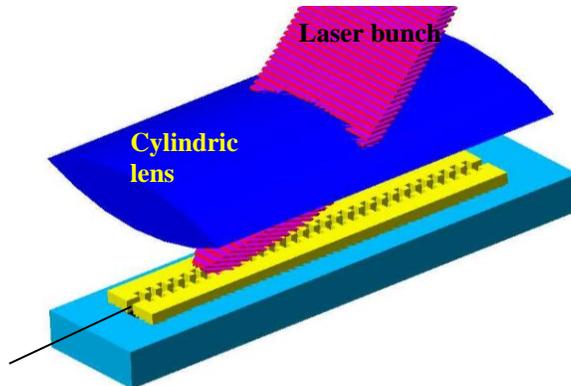

Figure 9: Accelerating structure is open at one side. The beam is going inside at distance ~ $\lambda_w/4$ from the bottom plate (magenta). Laser bunch is shown painted rose.

Parameters of accelerating complex are represented in Table 2 below, see [1, 2] for details.

Table 2: Parameters of Collider Driven by TLF

| | |
|---:|:---|
| Energy of $e^{\pm}$ beam | $2 \times 1\ TeV$ |
| Total two-linac length | $2 \times 100\ m$ |
| Wavelength of laser radiation | $\lambda_{ac} \cong 1\mu m$ |
| Luminosity | $10^{35} cm^{-2} s^{-1}$ |
| Main linac gradient | $>10\ GeV/m$ |
| Bunch population, $N_b$ | $3\ 10^5$ |
| Bunch length $\sigma_y$ | $0.1\ \lambda_{ac}$ |
| $\sigma_x$, $\sigma_z$ | $0.025\ \lambda_{ac}$, $0.1\ \lambda_{ac}$ |
| $\gamma\varepsilon_x / \gamma\varepsilon_y$ | $5\cdot 10^{-9}/1\cdot 10^{-9}\ cm\text{-}rad$ |
| No. of bunches/train | $<30$ |
| Laser flash energy | $100\ \mu J/section$ |
| Total laser flash energy | $2 \times 0.3 J$ |
| Laser beam power | $2 \times (0.3\text{-}3) kW$ |
| Laser density@Acc.Structure | $0.3\ J/cm^2$ |
| Illumination time | $<0.3\ ps$ |
| Length of section | $\approx 3 cm$ |
| Repetition rate | $1\text{-}10\ kHz$ |
| Damping ring energy | $2\ GeV$ |
| Wall plug power* | $2 \times (3\text{-}30)\ kW$ |

\* Without supplementary electronics.

With this system it is possible to arrange collisions with $e^{\pm}, \mu^{\pm}, p^{\pm}$ and even $\pi^{\pm}$ [2] in any combination of these.

**MODEL FOR FlexPDE**

For calculation of fields of radiation we used a FlexPDE code [7]. Model of acceleration structure with 15 cells was erected for this modeling. The front and back walls were made resistive. Calculation of wakes was an issue. Equations for vector and scalar potentials were appointed for modeling

$$\Delta \vec{A} - \tfrac{1}{c^2}\ddot{\vec{A}} - \kappa\dot{\vec{A}} = -\mu_0 \vec{j}, \quad \Delta U - \tfrac{1}{c^2}\ddot{U} = -\tfrac{1}{\varepsilon_0}\rho(x,y,z,t), \tag{9}$$

where $\mu_0, \varepsilon_0$ are magnetic permeability and electric permittivity of vacuum respectively, $c$ is a speed of light. The term with decrement $\kappa\dot{\vec{A}}$ describes the losses in walls-$\kappa = 1/(c^2\tau) = 1/(Q\lambda_{ac}c)$, $\tau$ is the field decay time, $Q$ is a quality factor. Such introduction of losses suggests volume losses although the real losses happen on surface. The volume losses introduced by such a way are self-consistent. For the bunch shape an approximation was used as (motion is going in *y*-direction, $v=\beta c$)

$$\rho = 0, \text{if } y/\sigma_y < -\pi/2 + vt/\sigma_y \text{ or } y/\sigma_y > \pi/2 + vt/\sigma_y,$$

$$\text{else } \rho \cong \tfrac{2Q}{\pi^2 \sigma_x \sigma_y \sigma_z} Sin^2(\tfrac{y-vt}{\sigma_y} + \tfrac{\pi}{2}) \cdot \exp(-\tfrac{x^2}{2\sigma_x^2} - \tfrac{z^2}{2\sigma_z^2}).$$

Usage of sin-like-squared shape of bunch allows it's localization in 3D without long tails in longitudinal direction. In a transverse direction finite extension is less important.

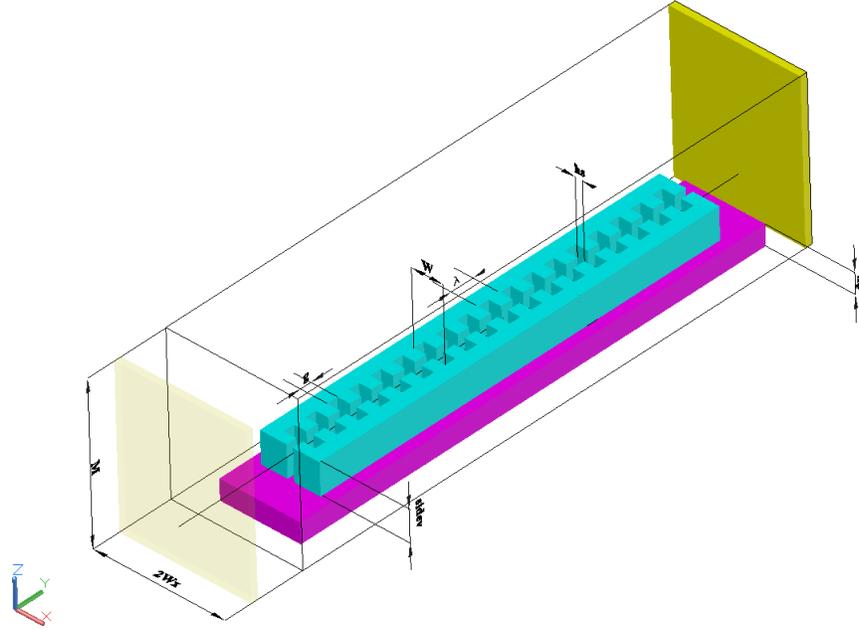

Figure 10: The geometry of problem box. Structure with 18 cells is shown in this Figure. In calculation a 15-cell structure was used for simplicity.

Boundary conditions for vector potential were chosen to be zero for tangent to surface components $A_\tau = 0$; normal components satisfy conditions $\partial A_n / \partial n = 0$, potential $U = 0$ at boundary. Current density, magnetic and electric fields defined as usual

$$\vec{j} = \rho\vec{v}, \quad \vec{B} = curl\vec{A}, \quad \vec{E} = -\partial\vec{A}/\partial t - gradU . \qquad (10)$$

The current has only *y*-component basically. Energy lost by bunch evaluated by

$$\Delta\varepsilon = \int dt \int (\vec{j}\cdot\vec{E})dV = -\int dt \int \vec{j}\cdot(\partial\vec{A}/\partial t + gradU)dV . \qquad (11)$$

Parameters of structure represented in Table 3.

Table 3. Structure Dimensions

| | |
|---|---|
| Wavelength, $\lambda_{ac}$ | 1μm |
| Width of cell, $W$ | 0.6 $\lambda_{ac}$ |
| Accelerating gap, $g$ | 0.5 $\lambda_{ac}$ |
| Height of structure, $h$ | 0.8 $\lambda_{ac}$ |
| Width of pass slot, $\delta$ | 0.2 $\lambda_{ac}$ |
| Qality factor, $Q$ | 10 |

(While making measurements of characteristics of structure in presence of air, one should take into account dielectric permittivity of air, 1.000536·$\varepsilon_0$).

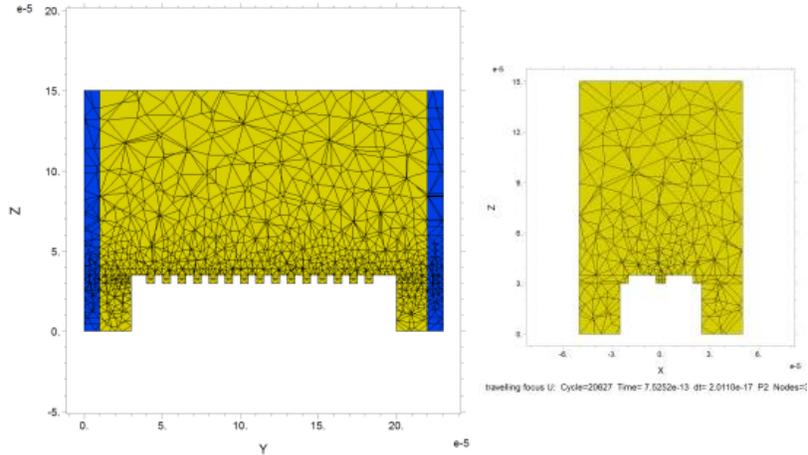

Figure 11: Mesh. The side regions with increased dissipations are painted blue.

Moving mesh generated automatically for delivering accuracy of calculations $\cong 5\cdot 10^{-4}$, see Fig.11. Calculated wake-fields are represented in Figs. 12-10. Basically each of these figures is a frame from movie created by code.

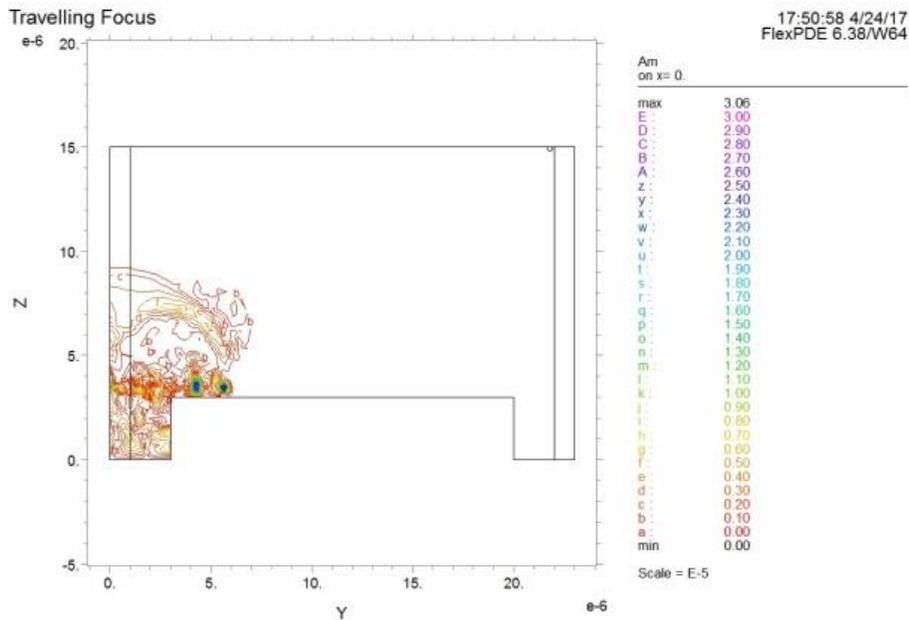

Figure 12: Wake fields contour in a plane running across the center of foxhole structure. Bunch passed two cells.

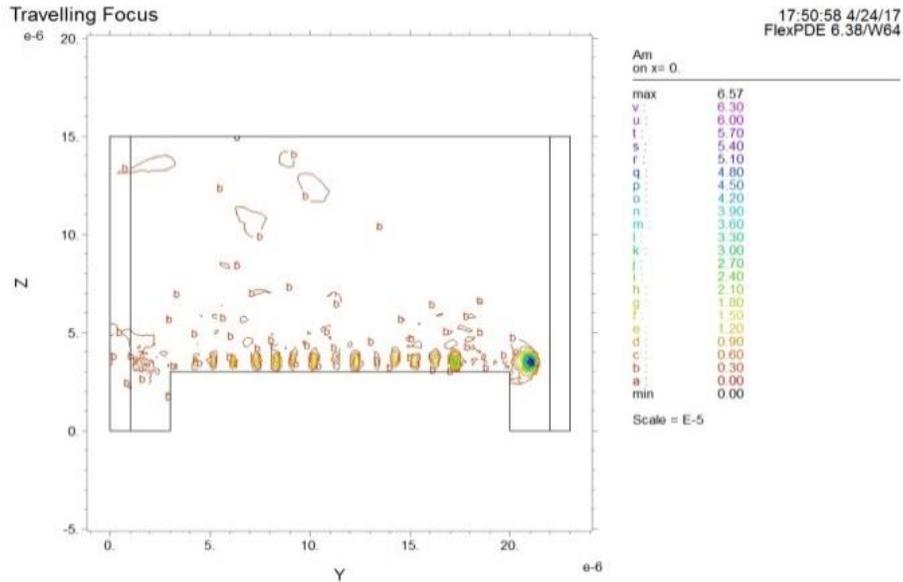

Figure 13: Module of vector potential for the bunch, exited the structure. As the plane of figure is running along the mid of structure, the walls of structure do not appear.

One can see on Fig.13, that the fields are practically scattered in outer volume to the moment of arrival of bunch to the boundary; they are located in cells mostly.

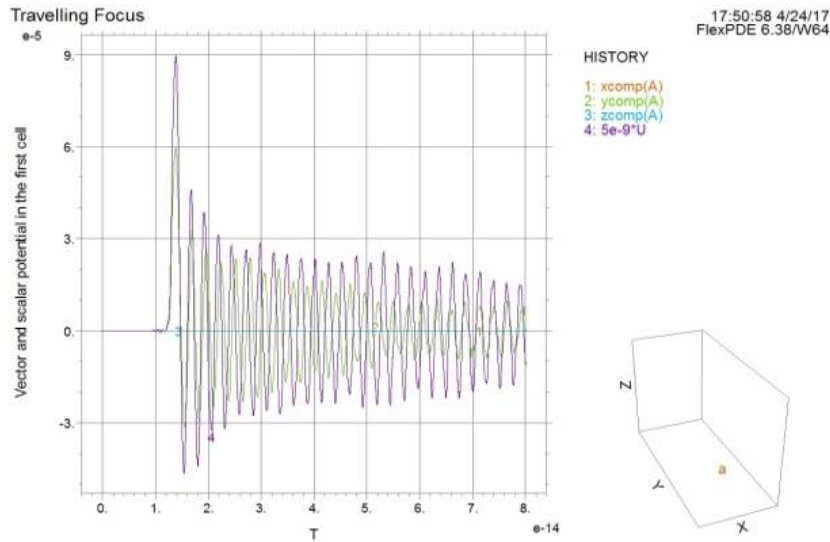

Figure 14: The fields in the first cell as functions of time.

Resonant frequency of cell could be extracted from this graph in Fig.14. This procedure will take into account existence of slots *etc.,* and it could be recommended for any complicated structure as well. Algorithm is straightforward.

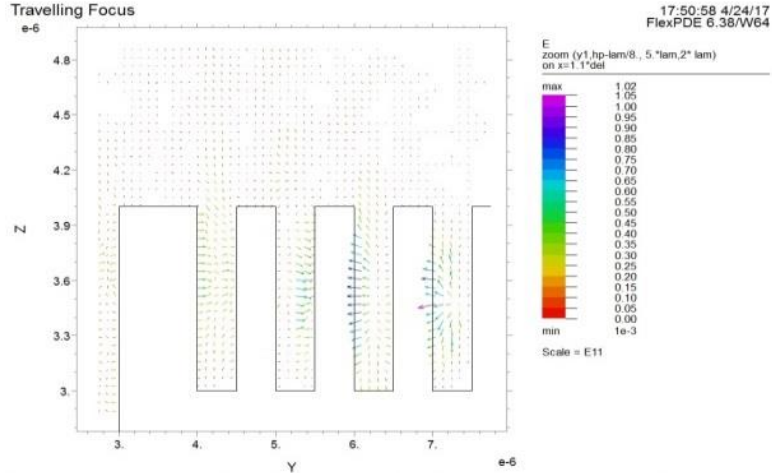

Figure 15: Vectors of $\vec{E}$, while the bunch is in a cell.

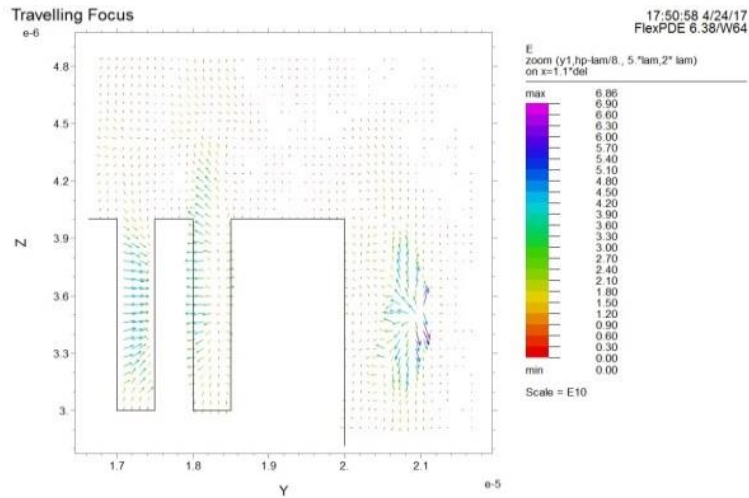

Figure 16: Vectors of *E*, while the bunch came out of structure. Moving window; a frame from a movie. The plane of cut is shifted from the center of structure, so the cells are visible here.

Calculations take ~4.5 hours with PC equipped with Xeon E5-1620 for structure with 15 cells with accuracy ~$10^{-3}$ and 12 individually recorded movies.

We would like to underline that as the wave fronts reach the open side of structure practically parallel to it, the periodicity of cells corresponds to $\pi$ mode as a simplest one for this geometry. Other than $\pi$ mode regime will require some tilt of wave fronts or cells spacing [2].

One should remember also that scaling of wakes in a cell is going linearly with dimensions. As the distances between focusing elements (see [2]) are also dropping linearly with wavelength, so the wakes action remains about the same as for X-band structure. One should remember that collider with TLF requires ~ $3 \times 10^5$ particles only.

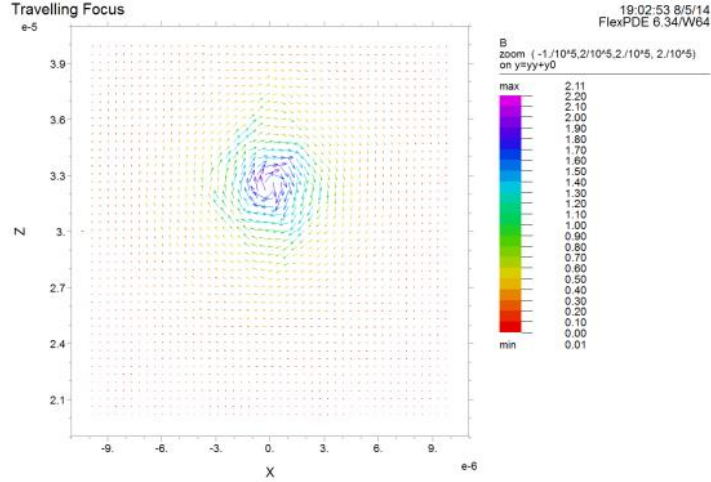

Figure 17: Vector $\vec{B}$ across the center of bunch which escaped the structure. A frame from a movie.

## SUMMARY

In this publication we concentrated on the wake-fields calculation in a foxhole-type structure. The code written in FlexPDE demonstrates stable operation and results of calculations look realistic. Calculations show, that the field induced by the bunch in each individual cell exists for few periods, demonstrating agreement with quality factor $Q=10$. This is indication of fact that induced radiation does not escape easily from cells for chosen coupling. This was found first in [2] with GdfidL code [8].

Choice of wavelength $\lambda_{ac} \approx 1\mu m$ makes fabrication more difficult, than for $\lambda_{ac} \sim 10\mu m$ one. From the other hand operation with shorter wavelength makes laser system easier. Final choice could be done later.

Scalar potential $U$ demonstrates a wave-like behavior, as it could be seen from Fig.14. Basically this could be supported thanks to Lorentz calibration which is hidden behind equation (9): $div\vec{A} + 1/c^2 \cdot \partial U/\partial t = 0$.

Tiny transverse dimensions of passing slots between cells $\delta = 0.1\lambda_{ac}$, require emittances $\gamma\varepsilon_{x,z} \approx 10^{-9} cm \cdot rad$. These emittances could be obtained for the bunch with population $N_b \approx 3 \cdot 10^5$ by scrapping extra particles from bunches populated traditionally- with $N_b \approx 2 \cdot 10^{10}$, but having emittances $\gamma\varepsilon_{x,z} \approx 10^{-7} cm \cdot rad$, see [2]. (ILC damping ring vertical emittance corresponds to the horizontal emittance in TLF system).

Wakes generated by the bunch mostly at the edges of structure, so smoothing of these regions require attention.

---

[4] Electronic version is available at http://www.lns.cornell.edu/public/CBN/2005/CBN05-6/cbn05-6.pdf.
[5] Electronic version available at http://www.lns.cornell.edu/public/CBN/2004/CBN04-6/phys_found.pdf